# A Middleware for The Internet of Things

Mahmoud Elkhodr, Seyed Shahrestani and Hon Cheung

School of Computing, Engineering and Mathematics, Western Sydney University, Sydney, Australia

## Abstract

*The Internet of Things (IoT) connects everyday objects including a vast array of sensors, actuators, and smart devices, referred to as "things" to the Internet, in an intelligent and pervasive fashion. This connectivity gives rise to the possibility of using the tracking capabilities of things to impinge on the location privacy of users. Most of the existing management and location privacy protection solutions do not consider the low-cost and low-power requirements of things; or, they do not account for the heterogeneity, scalability, or autonomy of communications supported in the IoT. Moreover, these traditional solutions do not consider the case where a user wishes to control the granularity of the disclosed information based on the context of their use (e.g. based on the time or the current location of the user). To fill this gap, a middleware, referred to as the Internet of Things Management Platform (IoT-MP) is proposed in this paper.*

## 1. Introduction

The Internet of Things (IoT) is the future of the Internet. It provides societies, communities, governments, and individuals with the opportunity to obtain services over the Internet wherever they are and whenever they want. The IoT enhances communications on the Internet between not only people but also things. It introduces a new concept of communication which extends the existent interactions between humans and computer applications to things. Things are objects of the physical world referred to as physical things, or the information world referred to as virtual things [1]. Things are capable of being identified and integrated into the communication networks. Physical things such as industrial robots, consumer products, and electrical equipment, are capable of being sensed, actuated and connected to the Internet. More specifically, a physical thing can be described as a physical object equipped with a device that provides the capability of connecting to the Internet. The International Telecommunication Union (ITU) defines a device in the IoT as a piece of equipment with the mandatory capabilities of communications, and the optional advanced capabilities of sensing and actuating [1]. On the other hand, virtual things are not necessarily physical or tangible objects. They can exist without any association with a physical object. Examples of virtual things are multimedia contents [2] and web services which are capable of being stored, processed, shared and accessed over the Internet. A virtual thing may be used as a representation of a physical thing as well, such as the use of objects or classes in object-oriented programming approaches [3].

Communications in the IoT can occur between not only the users and things, but also exclusively between things. These include communications between physical things, (also known as Machine-to-Machine communications), between virtual things, as well as between physical and virtual things. This heterogeneity of communications extends computation and connectivity in the Internet to anything, anyplace and anytime. As a result, the IoT is expected to be used in numerous application domains, including but not limited to, manufacturing [4], smart cities [5], agriculture and breeding [6], environmental management [7], and smart homes [8]. Significantly, the IoT enables the sharing of information between different domains [9]. For instance, in the healthcare sector, the IoT supports the sharing of medical information between various healthcare professionals and therefore it enhances the delivery of health services [10]. From a networking perspective, the IoT can be described as a heterogeneous network that connects together many wired and wireless networks, including low-power wireless networks and personal area networks, with an increasingly complex structure. This heterogeneous network connects a range of devices together. It encompasses devices which connect to the Internet using various types of wireless, mobile and LAN technologies such as Wi-Fi, ZigBee, Bluetooth, and 3G or 4G technologies among other evolving communication technologies.

The IoT has the potential to provide an intelligent platform for the collaborations of distributed things via local-area wireless and wired networks, and/or via a wide-area of heterogeneous and interconnected networks such as the Internet [11]. The availability of information coming from non-traditional computer devices in the IoT will change society and transform businesses. In 2010, the IoT market value was estimated to be worth more than 100 billion dollars by 2020 [12]. In 2013, Cisco forecasted that the economic value created by the IoT will exceed 14.4 trillion dollars in 2020 [13]. Cisco increased its forecast in 2014 to 19 trillion dollars [14]. Furthermore, IC Insights predicts that the number of new connections to the IoT will grow from 1.7 billion devices in 2015 to more than 3.1 billion devices in 2019 [15]. On the other hand, Cisco estimates that the number of connected devices to the Internet will exceed 50 billion in 2020 [16]. BI Intelligence also predicts that the number of things connected to the Internet will grow by 35% between the years of 2014 and 2019 [17]. Consequently, these forecasts and predications highlight the significance and economic value of the IoT, and the role it plays in elevating communications on the Internet.

Beyond the massive technological opportunities and benefits of the IoT, important challenges such as interoperability, security, and privacy arise [18]. However, the complexity in addressing the IoT challenges lays in the fact that these challenges are correlated together [19]. That is, there is a need to achieve full interoperability among the various types of things that communicate, seamlessly, over heterogeneous communication networks. This interoperability needs to be achieved while guaranteeing the best possible Quality of Service (QoS) and highest degree of security, trust, confidentiality, and privacy. Additionally, the IoT presents unique challenges for energy-efficient operations [20]. Many things in the IoT need to run for years over batteries [21]. Therefore, until contemporary power sources or energy harvesting solutions are developed, energy consumption remains a challenging issue in the IoT.

Therefore, the potentially massive number of things, their diversity, and the seamless and heterogeneous nature of communications encountered in the IoT create many prominent challenges in terms of management, interoperability, security, and privacy. Although some people may willingly reveal their location information in order to obtain location-based services, few would be comfortable having their locations constantly collected by the billions of things envisioned in the IoT. The diffusion of wireless communication networks and the technical advancements of location positioning techniques in the IoT provide IoT applications with the capabilities of automatically sensing and actuating the things' environments, communicating, and processing the information collected by other things which may reveal their owners' locations, with a high degree of spatial and temporal precision. Thus, the flow of information and actuation events in the IoT comprises the exchange of personal and contextual information supplied by things, including location information. This gives rise to the possibility of using the tracking capabilities of things to violate the location privacy of users.

International Journal of Computer Networks & Communications (IJCNC) Vol.8, No.2, March 2016

Currently, many studies are involved in developing solutions to solve the various research challenges facing the IoT. Specifically, the development of new middleware solutions for the IoT is an active area of research. For instance, the platform developed by Axeda [22] provides a Cloud-based system for managing things connected to the cloud. It provides security services for securing the communications between things and the cloud system as well. Other solutions such as the Kaa platform [23] also offers a middleware solution to connect things to the cloud. However, Kaa requires the integration of a specific microchip in the hardware of the IoT device. Our work goes one step further by providing the users with location privacy protection capabilities in the IoT. The middleware, proposed in this work, is referred to as the Internet of Things Management Platform (IoT-MP). It provides users with fine-grained control over the granularity and disclosure settings of the users' location information in the IoT. It is based on a distributed architecture that utilises an agent, a manager, and a manager of managers paradigm. The IoT-MP is specially designed to accommodate the requirements of constrained things, and the heterogeneous and scalability characteristics of the IoT. It adopts an extensible design where things are virtually represented as attributes in a management database located at the manager. In this way, IoT applications can access things transparently over the Internet, irrespective of the underlying used communication technologies.

## 2. ARCHITECTURE OF THE IoT-MP

The architecture of the IoT-MP ranges from a simple two-tier architecture consisting of a manager and its associated agents to a distributed architecture which consists of agents, managers and a Manger of Mangers (MoMs). In the simple two-tier architecture, as shown in Figure 1, the IoT-MP adapts the structure used by traditional network management approaches. Here, the IoT-MP architecture consists of an agent that resides on a thing and a manager that manages things. The agent acts as a communication agent. It has the responsibility of transporting the data generated or collected by things to the manager. The term Managed Things (MT) is used throughout this work to refer to an agent and the device as one entity. The manager manages many MTs in a system. It is a one-to-many relationship between the manager and its associated MTs. The manager stores the received data, sent by managed things, in a database. The IoT-MP provides IoT and management applications, running on top of the manager, access to the data generated by things and stored in the database remotely over the Internet. The IoT-MP also provides IoT applications with a mechanism to send instructions to MTs e.g. sending actuation instructions to an MT.

In the distributed architecture, a Manager of Manger (MoMs) is introduced. This architecture is provided in Figure 2. The MoMs is a web application which contains an API, referred to as the management API, which allows the MoMs to communicate with many managers. The MoMs has also a database which stores the addresses of MTs. The MoMs' database is hierarchical (tree-structured), and each entry relevant to a managed thing is addressed through the Manager unique ID. Thus, the MoMs' database does not store any information collected from managed things. It only stores their addressing information i.e. through which manager an MT can be accessed. An IoT application running on top of the MoMs can provide services based on combining data collected from various managed things. Therefore, IoT applications and services are built on top of the M2M by supporting communications between things-to-things via their respective managers.

Consequently, the architecture of the IoT-MP provides a centralized model of several agents and a manager allowing for central management of things over local area network. On the other hand, the distributed architecture of several managers and a MoMs creates a distributed system where things communicate in a cooperative fashion rather than stand-alone manner. This flexibility in design caters for the specific communications requirements in the IoT. Given that most IoT devices may play different roles in both centralized and distributed operations setups.

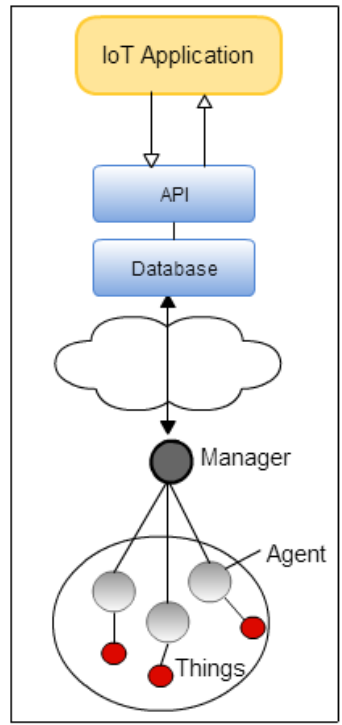

Figure 1- The IoT-MP Two-Tiers Architecture

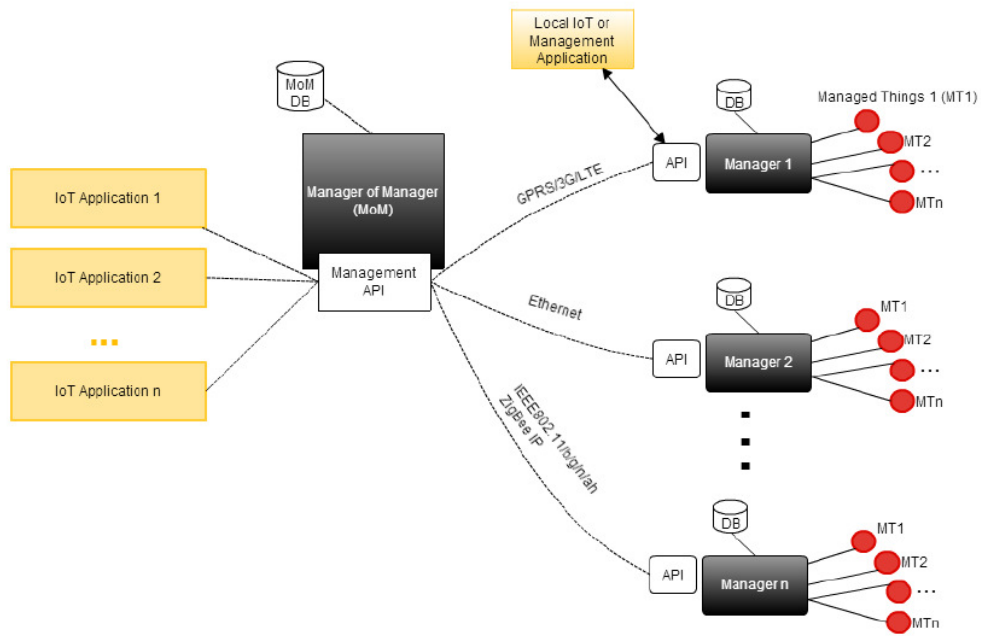

Figure 2- The IoT-MP Distributed Architecture

162

## 3. THE IOT-MP ENTITIES

In this section, the various entities of the IoT-MP are introduced.

### 3.1 Things, Agents, and Managed Things

Things are virtually represented using attributed on the management database of the IoT-MP. The management database is discussed in more details in Section 3.3. The representation of things using attributes is shown in Figure 3.

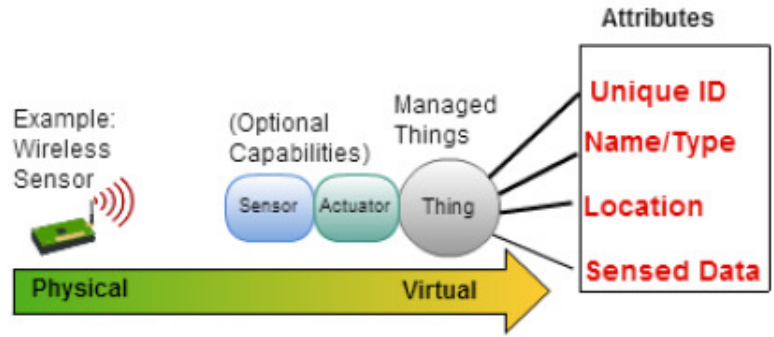

Figure 3- Things' Attributes

More specifically, things are represented using two types of attributes. The first group of attributes is referred to as "management attributes". They are used for the virtual representation of things. The second type of attributes is referred to as "behavioural attributes". These attributes are used to hold the data things sense or collect. They are also used to hold information relating to actuation events. The definition of these attributes provides a way to store information about things and the information they collect on the management database. Table 1 provides an example of management and behavioural attributes which represent a wireless sensor device.

Table 1 - Behavioural and Management Attributes example

| **Management Attributes** | **Behavioral attributes** |
|---|---|
| • ID | • Temperature |
| • Name | • Motion |
| • Serial Number | • Sound |
| • Firmware version | • Pressure |
| • IP, MAC address, network name or others | • Water detection |
| | • Fire detection |
| • Battery life | • Location (if mobile) |
| • Location (if fixed) | |

From Table 1, the management attributes are descriptors of a wireless sensor device. For example, the ID is used to uniquely identify a thing. The name, serial number, firmware version and the rest of these management attributes are also used as descriptors for things. These attributes are optional, and the decision whether to use them is left to the specification of things. For instance, not all things necessary have a firmware version. Thus, a wireless sensor device may elect to use only the ID and location, as an example, from the list of available management attributes. While, a smart enabled Wi-Fi device may use more management attributes. Similarly, the behavioural attributes are used as records in the management database. These are also optional as they are inheritably associated with the features and the applications of things. For instance, a temperature sensor that has the responsibility for collecting motion and temperature data from a given environment will choose the "Temperature" and "Motion" from the behavioural attributes lists. Other things with a different functionality or scope such as a fire detection sensor will choose the "Fire Detection" from the list of the behavioural attributes. Operators of things may also create new attributes in the management database to represent them.

On the other hand, as mentioned before, Managed Thing (MT) is a term used to refer to an entity which comprises of a thing and its agent. Figure 4 illustrates the architecture of an MT. It further shows that an agent has two main modules. The communication module is responsible for the communications with the manager. The second is the agent services module that is responsible for the execution of services such as generating alerts or processing an actuation instruction. Thus, the agent handles the communications between the manager and things by forwarding the collected data to the manager and by processing the requests sent by the manager. In summary, the agent has the followings responsibilities:

- Establish a communication with the manager.
- Sending data updates from MTs to the manager.
- Handling requests and their correspondent responses from/to managers.
- Receiving actuation instructions from the manager and passing them to the MTs
- Sends notifications to managers

To support the communications between MTs and their manager, a message exchange scheme is defined in the manager's communication module.

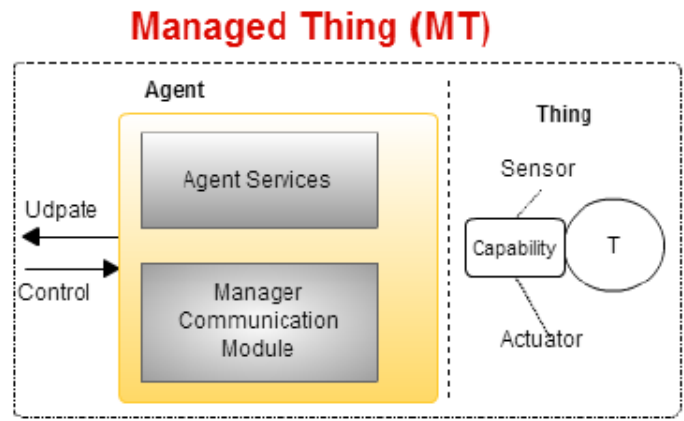

Figure 4 - Managed Things

## 3.2 Managers

A manager is an application that performs the operational roles of generating requests to retrieve information from managed things. A manager can also send actuation requests on particular MTs, which have actuation capabilities. A manager receives event-based notification reports on MTs generated by their agents as well. Additionally, a manager issues requests for management operations on behalf of an administrator or IoT application and receives notifications from agents. A manager maintains a management database which stores information about MTs. The manager accesses MTs' data stored in this database. The manager supports a management API based on the Restful architecture. This API allows IoT applications to request over the Internet information about MTs, which is stored in the management database. The API also allows management applications to monitor and control MTs. The management API is also used to communicate topology information of the network to the Manager of Managers (MoMs) over the Internet. The MoMs is a higher entity sitting on top of the manager. IoT applications run on top of the MoMs and send requests to access data of MTs under the manager supervision. Figure 5 shows how the manager interacts with things via agents.

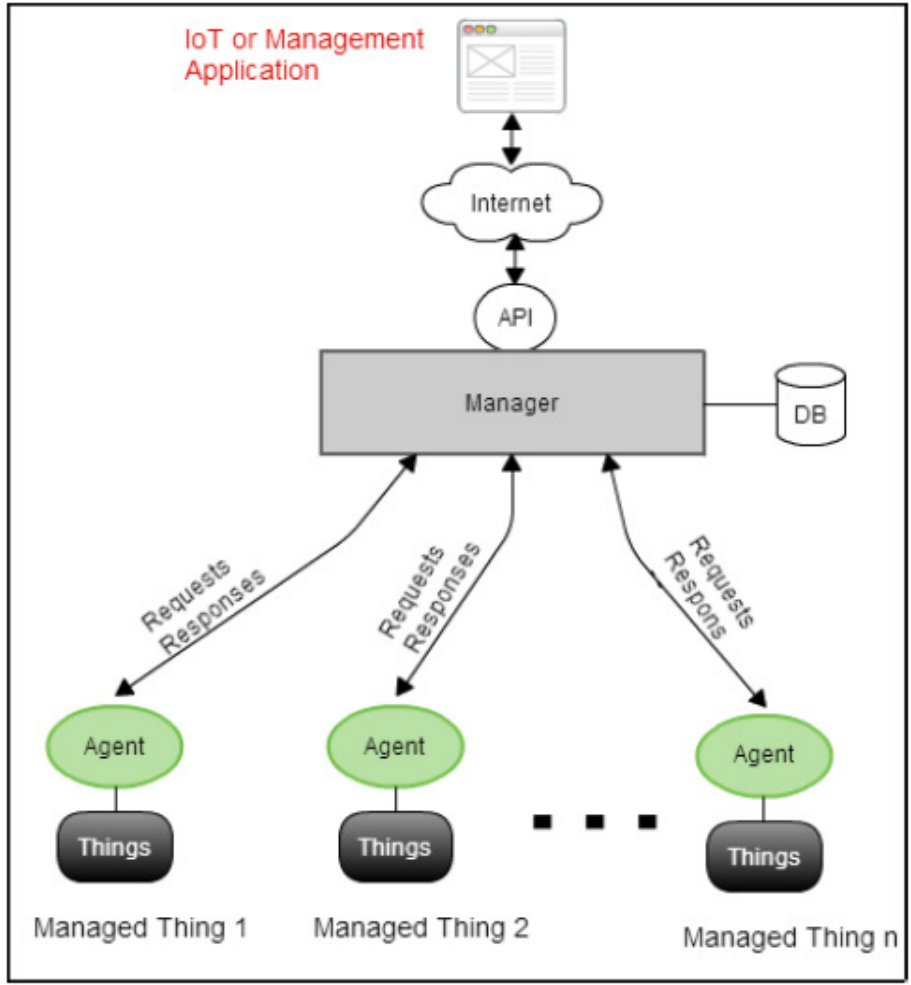

Figure 5 – Communications between the manager and MTs

Furthermore, the manager comprises of several modules which are used to provide management, security, and privacy capabilities among other operational services. These modules are introduced in Section 4.

### 3.3 The Management Database

The Management Database is illustrated in Figure 6. MTs' information is indexed in the database using the unique ID of the MT. The managed attributes describe the MT's type, name and other descriptors that can be defined by the administrator. The Behavioural attributes are used to store the data received by the manager from an MT.

165

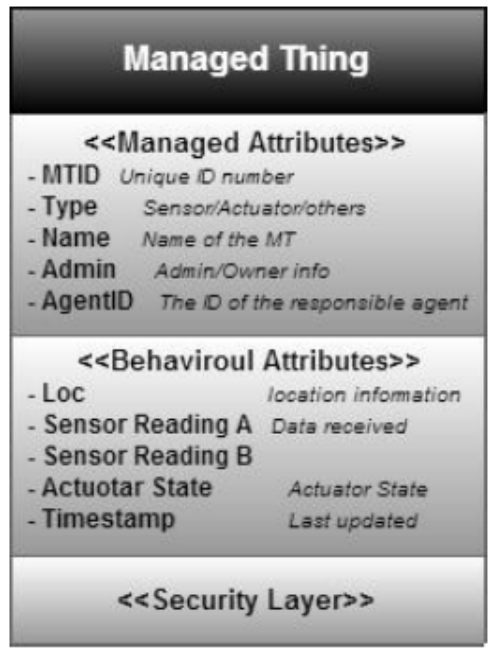

Figure 6 - MTs Entries

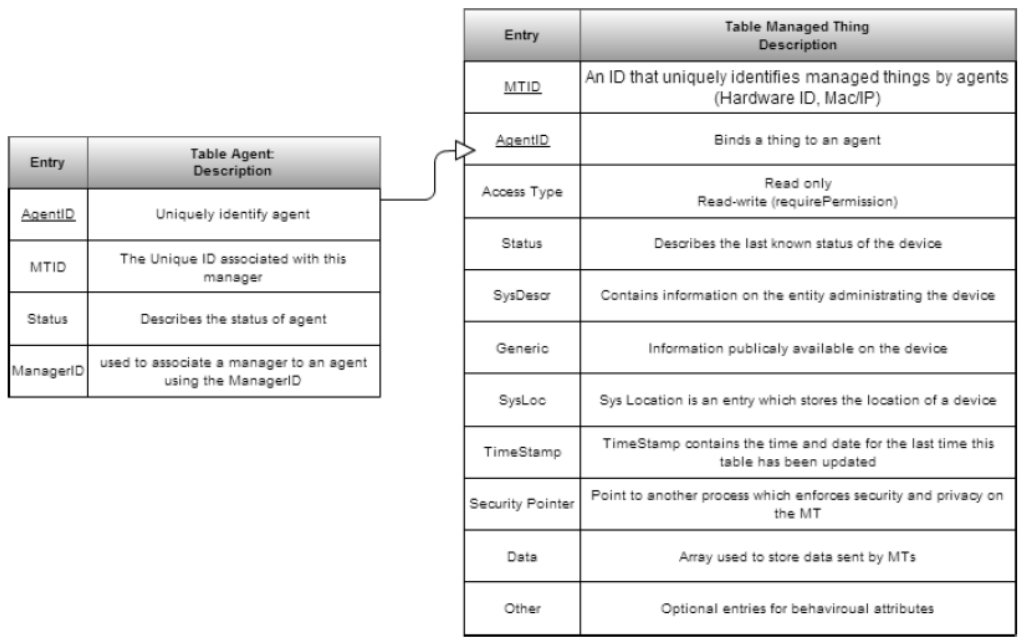

Figure 7 - Database Schema

Figure 6 shows examples of Managed and Behavioural attributes. For instance, the MTID is used to uniquely identify an MT. The Type, Name, Admin, and AgentID are other examples of managed attributes. The Loc, Sensor Reading A, and B are examples of behavioural attributes

166

International Journal of Computer Networks & Communications (IJCNC) Vol.8, No.2, March 2016entries in the database. The security layer is a pointer to another process that provides security and location privacy protections. The main parts of the database schema are provided in Figure 7. The table Managed Things, from Figure 7, contains the followings entries:

- MTID is an identifier used to uniquely identify things in the IoT-MP. It is assumed that each IoT device (things) is assigned a unique identifier.
- AgentID is an identifier used to identify agents. This AgentID is assigned by the manager. The MTID and AgentID are used to identify an MT. In the case where the agent is residing on the MT, the MTID is used in place of the AgentID.
- The security pointer points to another process which enforces access control disclosures policies on the data of MTs.
- The descriptions of the rest of the entries in the database are provided in Figure 7 and are self-explanatory.

### 3.4 Manager of Managers (MoMs)

The Manager of Managers (MoMs) provides mapping and routing services over the IoT-MP. Its functionality is similar to a router which performs "traffic directing" functions on the Internet. The MoMs forwards messages between managers' networks and IoT applications over the Internet. An IoT application accesses the MoMs remotely over the Internet. It can request information about an MT belonging to a particular manager's network or supply information that can be used by other MTs or IoT applications.

The communications between a manager and the MoMs are over HTTPS and handled by their APIs. The MoMs maintains a hierarchical database which stores the addresses of MTs and their managers. The MoMs does not store in its database any operational data on MTs. Thus, similarly to a routing table, the MoMs database is used to store information about the topology of the network. That's which manager oversees which group of MTs. This information is used by the MoMs to determine where information should be directed within the IoT-MP network.

The MoMs database consists of a table which has three fields:

- The ManagerID: This is used to uniquely identify managers
- The IP address of the manager
- The list of MTID belonging to each manager network

The scenario is as follows: the MoMs receives a request to access a certain MT data from an IoT application over the Internet (HTTP request). The request must specify at least the MTID of the MT under request and the "AppID" of the application, along with some other credentials used for authentication. The MoMs then lookups the MoMs' database and retrieves the ManagerID and IP address. The MoMs forwards the request to the correspondent manager and waits for a response. Lastly, the manager responds to the MoMs with either an error message or with the data being requested.

In the rest of this work, instead of passing through the MoMs each time a request is issued by an IoT application, we will simply refer to the request as being issued by an IoT application to a manager.

## 4. THE ARCHITECTURE OF THE MANAGER AND ITS MODULES

The manager is designed in a modular format, so it is possible to extend its functionalities and capabilities. As shown in Figure 8, the manager has seven main modules. These are described in the subsections that follow this section.

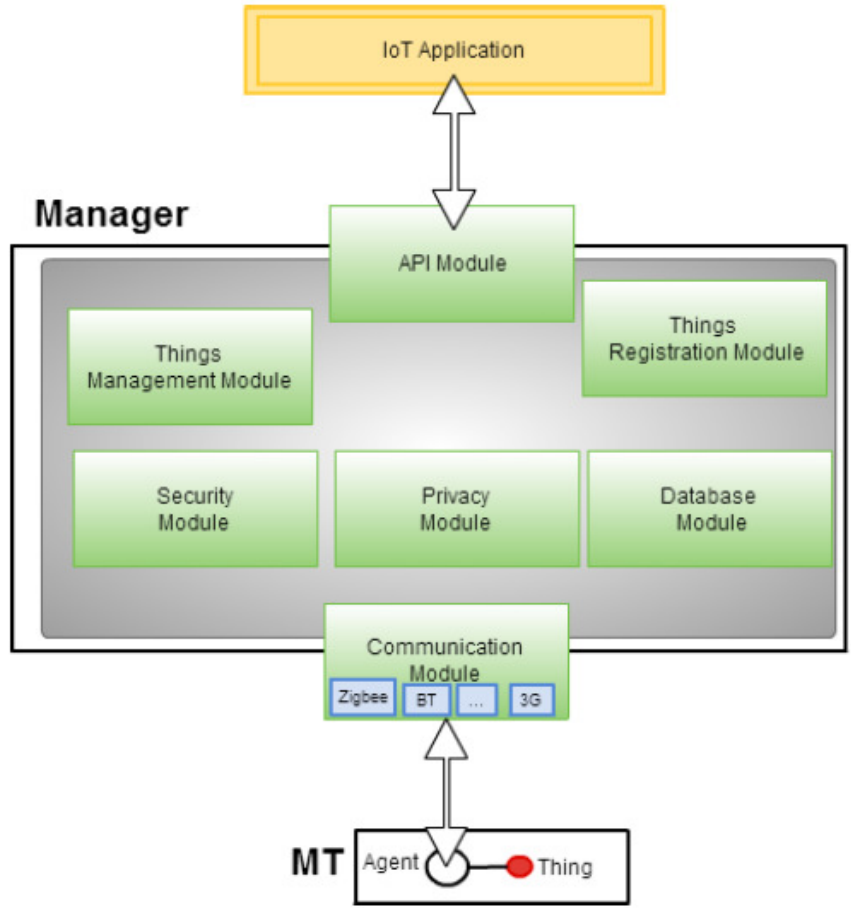

Figure 8- The Manger's Modules

### 4.1 The Communication Module (CM) and Things Management Module (TMM)

The communication module (CM) is responsible for handling the communications between the manager and the managed things. The CM supports the communications of two types of messages: management messages and operational messages. Management messages are handled by the management module and are used by the manager to obtain management information such as network states information, connection information, and dynamic performance information. The Operational messages are mainly used to get status information and updates from managed things. These messages are based on two messages "Get" and "Update". As their names suggest, the 'Get' messages are used by the manager to get information from an MT. The "Update" message is utilized by the MT to send an update to a manager.

The Things Management Module (TMM) defines the messages used by the manager to manage MTs. These management messages are adapted from SNMP. Mainly, the TMM adapts three SNMP messages: the SET, GET, and Alert SNMP messages. They are used to inspect and communicate information about MTs to the manager. The message exchange scheme supported by the TMM along with their messages' format has been previously introduced in [24].

### 4.2 The Things Registration Module (TRM)

The Thing Registration Module (TRM) is responsible for registering things on the IoT-MP. It involves the process of things bootstrapping, connecting and joining the network, and creating an entry for the thing in the database. A thing joins the IoT-MP network and becomes an MT by connecting to a manager using an agent. Thus, it is assumed for the communication to be established with a manager; a thing must rely on an agent that is capable of communicating with the manager. As previously mentioned in Section 3.1, the agent is a software that can be installed on the thing, or it could reside on another device in the thing's local area network. For example, a Bluetooth slave device may communicate with a Bluetooth Master device which implements the agent capability. In this example, the Bluetooth Master device is considered to be the agent. The combination of the master and slave device is referred to as the MT. It is assumed that communications between the Bluetooth Master and Slave device are already established. Therefore, assuming a thing already has an agent capable of communicating with the manager; a thing connects to the manager and becomes an MT using one of the followings three methods:

   I.     by joining the network directly;
   II.    by requesting to join the network through associations;
   III.   by reconnecting to the network.

The sequence diagrams for the method I and II are provided in Figure 9.

In the direct joining method I, the thing is pre-programed with the IP address of the manager. This allows the thing to send a DIRECT-JOIN request, via an agent, to the manager. In method II, the process of connecting an MT to the manager through association is similar to that of a Wi-Fi device which associates with an access point. Method III is used when an MT loses its connection with the manager and tries to re-connect. This can happen when an MT moves out of range of the local-area network or when the manager becomes unavailable. In such a case, the MT will try to reconnect and join the network using Method I or II. The difference between method I and II on one hand and that of III, on the other hand, is that in method III there is no need to create a new record in the database for the MT. Instead, the manager updates the existing database record.

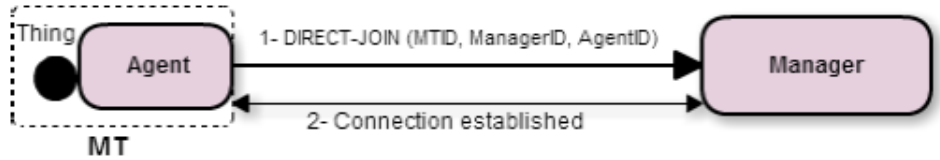

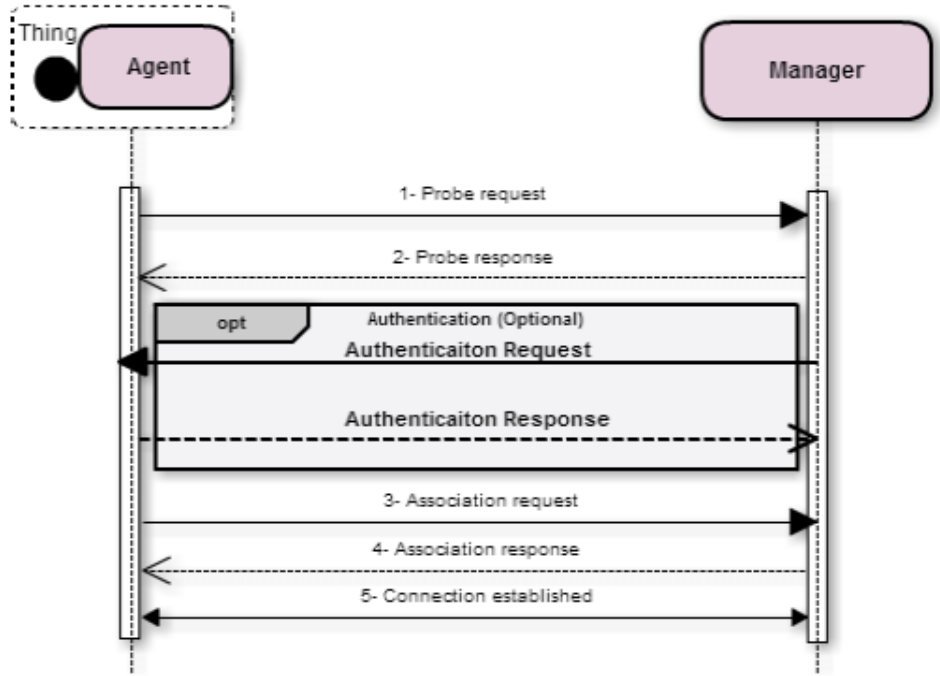

Figure 9 - Method I and II sequence diagrams

### 4.3 The Security Module (SM)

This module provides (1) security services between managed things and a manager and (2) between IoT applications and the manager using a security profile. In (1), the security module enforces that an agent must first be registered with the manager and approved by an administrator. The manager keeps a database of authorized agents using their AgentIDs. Therefore, the Security Module (SM) ensures that only authorised remote agents are allowed to connect to the manager. To achieve this, the SM implements these two directives:

a) The SM denies unknown remote agents from connecting to the manager.
b) Remote agents need to be manually approved by an administrator before they can communicate with the manager.

This work also assumes that the identity of an agent (AgentID) is unique and protected against classical identity threats such as identity forging, poison attack, and the likes. In order to relax the condition implied in b), which requires the manual approval of agents by a human and to automate the process, classical access control and identity management solutions can be employed to preserve the identity of agents. For instance, a digital certificate solution can be combined with an Attribute-based access control (ABAC) system to secure the process of establishing a security association between an agent and a manager. However, identity management is not the focus of this work. Hence, we assume that an agent is always securely connected to a manager. This includes encrypting the communication channel between these two entities.

In (2), IoT applications can connect to the manager by sending requests over HTTP to the management API. The role of SM is to provide optional security services which protect HTTP sessions using SSL. The SM also contains a security profile enabling owners or admins of MTs to define security disclosure policies. These policies are used to define whether the MTs' data stored in the management database can be accessed by a given entity (i.e. an IoT application). Thus, the security profile allows the owner of an MT to define to whom the information collected from MTs is disclosed. This is achieved by providing the users with a UI enabling them to maintain a list of authorized entities. Furthermore, the security profile is used to create policies which govern whether information collected from MTs should only be disclosed to IoT application over secured channels or no. An activity diagram showcasing the role of the SM and its security profile is provided in Figure 10.

From Figure 10, the manager receives a Get Request message from an IoT application requesting some information from an MT. The details of how these messages are processed are provided in the Section management API. Upon the receipt of the message "1.GetRequest", the manager proceeds first into checking whether the requesting entity (the IoT application in this case) is authorised to access the requested MT data. The manager does that by sending to the SM the message "2.1CheckPolicy" which returns a flag that can be either "True" which means the request is approved or otherwise "False".

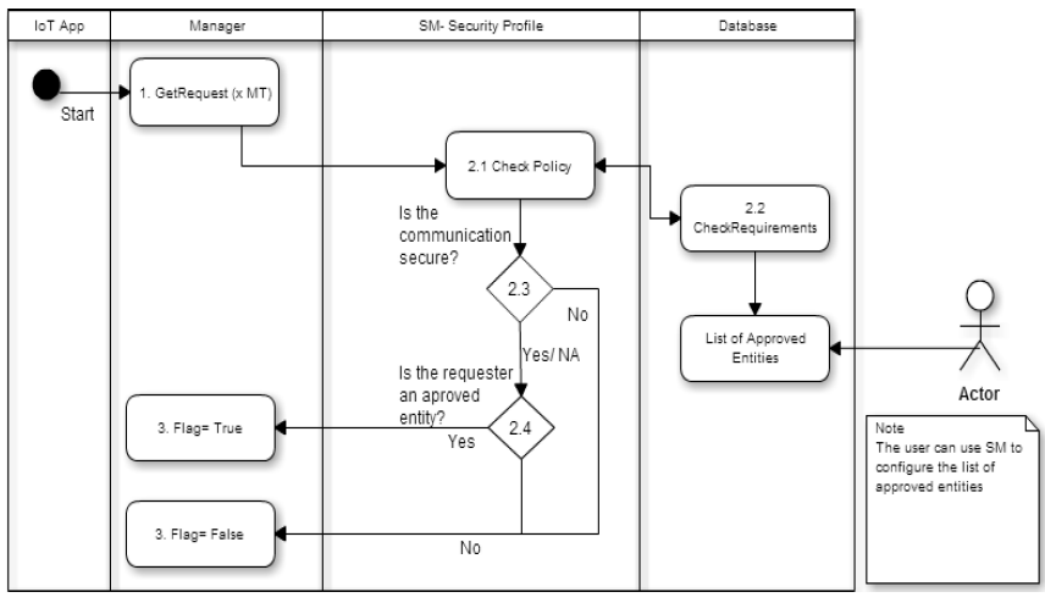

Figure 10- SM Activity Diagram

The message "2.1CheckPolicy" has the following sub-messages:

- It starts with the message "2.2 CheckRequirments" which retrieves from the database the security policies stored by the user that govern the disclosure of the information of the MT under request. Note that in Figure 10, the owner or admin of an MT has access to the list of approved entities. This enables him or her to add to the list an approved entity and to indicate whether their MT's data should only be disclosed over secured connections.
- The "Is communication secure" decision point checks whether the policy indicates that the MT's data under request should only be disclosed to authorized entities. If the request does not meet this condition, then the decision point takes the "No" path and sends a "False" return message back to the manager. If this criterion is met, then the "Yes" path moves to another decision point.
- The "is the requester approved" decision points checks whether the requester (the IoT application is this case) is on the list of the approved entities. If "Yes" then the flag is set to "True". In this case, the manager receives a return message "True" which indicates that the SM approves this request. Otherwise, if the result is "No" then the request will not be approved, and the original Get Request message will be replied with an error message. It should be noted that the manager does not disclose the MT's information in the case of "Yes" until the receipt of an approved message from the Privacy Module.

### 4.4 The API Module

The API module consists of an application programming interface (API), referred to as the Management API. This API provides an integrated and enhanced access interface for IoT applications. It enables IoT applications to participate in a communication in the IoT-MP. That is the API allows IoT applications to retrieves MT's data stored in the database remotely over the Internet. It also allows IoT applications to send actuation instructions over the Internet to relevant MTs. The API design is based on the Representational State Transfer (REST). It implements a set of messages known as HTTP verbs (GET, POST, PUT, and DELETE) as shown in Figure 11. REST is platforms and languages independent. Thus, it supports multiple platforms (HTML, Unix, Android, Windows, iOS, etc.) and does not require the use of a specific programming language (e.g. Java). Hence, theoretically, the API module can support a variety of IoT applications regardless of the platform or programming language they run or use. However, these applications need to support HTTP. Access to the management API is secured using HTTPS in combination with a token-based authentication process.

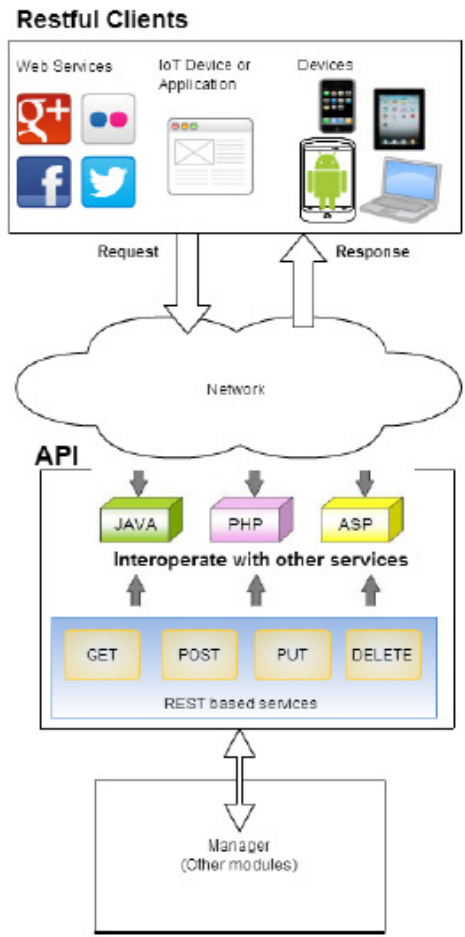

Figure 11 - The Restful based Management API

From Figure 11:

- **GET:** used by an application to request from the manager access to an MT's data which is stored in the management database.
- **POST:** used by an application to send actuation instructions to the manager.
- **PUT:** used for sending inserts and updates commands by a management application to the management database.
- **DELETE**: used by a management application to delete data from the management database.

PUT and DELETE commands are reserved for management applications only. Consequently, an IoT application can only use two commands: GET, which is used to retrieve information about MTs, and POST which is used to contribute with information or to send actuation events. Designing APIs for the Internet of Things is an active area of research. The area of API design has recently advanced to combine authorization with authentication in its design. For instance, JSON web tokens can be used with OAuth 2.0 providing identity delegation services as well. The API designed in this work is intentionally kept simple. The aim is to simply provide a valid method to authenticate an IoT application to the IoT-MP and thus to allow this IoT application to engage in a communication on the IoT-MP. Other researchers or future works may look into designing an optimized API for the IoT-MP.

### 4.4.1 Authorizing IoT Applications

The IoT-MP requires that IoT applications must be authenticated first by the manager. That is, for an IoT application to send an actuation instruction, contribute with some data, or request access to an MT's information over HTTP, the IoT application must be first authenticated by the management API. This authentication is done by communicating certain credentials to the management API. The authentication process is as follows:

- The administrator or owner of the IoT application must first register the IoT application on the IoT platform using a unique identifier referred to as the "AppID". This "AppID" can be chosen by the person registering the IoT application. However, the API must verify that the AppID is unique. That is, the IoT-MP will verify that no other IoT applications, registered on the IoT-MP, are using the same AppID. Alternatively, an IoT application can utilize a unique AppID automatically generated by the API registration interface.
- Upon successful registration, the IoT application will be issued with a unique "Secret Token".

For an IoT application to be granted permission to access MT's data, the IoT application must authenticate itself by sending a valid access token in the HTTP request header to the manager along with its AppID. The access token is, in fact, the Secret Token which was assigned to the application upon registration. The supplied credentials will then be used in the authentication flows between an IoT application and a manager.

The authentication flow starts with the manager validating the Secret Token initially passed in the request. After the Secret Token is validated, the manager uses it to establish a security context for the IoT application. That is, if the IoT application is successfully authenticated, then the request moves to another stage of authorization conducted by the Security Module. Recall, the SM has to responsibility for checking whether an IoT application is on the list of approved entities by the owner of the MT under request. This process is done using the function CheckPolicy() from the security profile as previously described in Section 4.3. Unauthenticated requests issued by the IoT application will be replied with an *HTTP 401 Unauthorized* response message and will not reach the authorization stage.

The design of the API is based on the JSON web API. The main parts of the API design are summarized to as follows: The format of the JSON based web token is three strings separated by a dot (.). Example: stringA.stringB.stringC. StringA is the header. StringB is the payload and stringC stores the signature. The header specifies the type of the request and the hashing algorithm in use which is SHA256. The payload carries the "AppID" and "Secret Token". The last part, stringC, is the signature. This is a hash function that combines the header, the payload, and a secret. The secret is the signature held by the manager. It allows the manager to verify existing tokens and sign new ones.

### 4.5. The Privacy Module (PM)

The Privacy Module (PM) combines three main components: the content analysis, privacy manager and semantic obfuscation components. These components were previously introduced in [25]. In brief, the PM provides the administrator of an MT with an added layer of location privacy protection. It enables the user to define privacy-disclosure policies. These policies are used by the manager to determine whether "to disclose" or "not disclose" the location of an MT when an IoT application request access to it. Additionally, it allows the user to control the granularity of the location information of an MT by attaching an obfuscation level to a policy. The architecture of the PM module is provided in Figure 12.

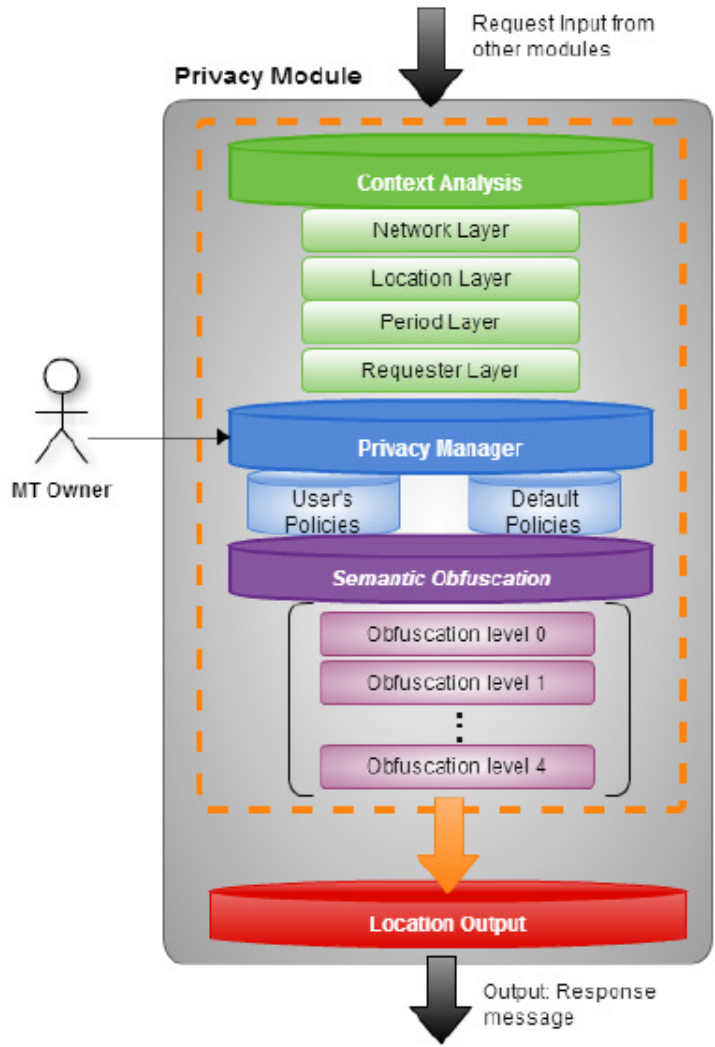

Figure 12 - Privacy Module

The specification and implementations of the context analysis and privacy manager components were previously introduced in [25] and that of the Semantic obfuscation component in [26]. The Privacy Module, through its three components, provides a context-adaptive method which enables the user to manage the location information disclosed by things based on a context-aware and policy enforcement mechanism. This mechanism takes into account both the user's informed consent and preferences. It provides the user with fine-grained control over the location information exchanged by things in situations where decisions about disclosing this information need to be made with minimal human intervention. The privacy preserving capabilities of the IoT-MP are also supported by a novel obfuscation technique, referred to as the Semantic Obfuscation Technique (S-Obfuscation). The S-Obfuscation enhances the performance of two classical obfuscation techniques by relying on a geographic knowledge when producing obscured location as reported in [26]. Therefore, the PM provides the user with the granule control over the location privacy of their MTs including to whom, when and in which context the location of their MT is revealed.

## 5. CONCLUSIONS

The development of solutions to support location privacy preservations is key factors for the proliferation of the IoT. Various privacy protection methods have been proposed in the literature to deal with the location privacy issue. However, most of these methods were designed to work with computers or mobile phones. They do not consider the low-cost and low-power requirements of things, or the heterogeneity, scalability, and autonomy of communications supported in the IoT. To address these shortcomings, in this paper, a middleware solution is proposed. It enables the management and preservation of location privacy of things in the IoT. The middleware accounts for the unique characteristics of things such as being lightweight, mobile across many heterogeneous domains and networks, and involved in seamless communications with other things or IoT applications. It encompasses a context-adaptive approach which enables the user to manage the location information disclosed by things based on a context-aware and policy enforcement mechanism. This mechanism takes into account both the user's informed consent and preferences.

The IoT-MP has a distributed architecture consisting of agents, managers and a Manager of Managers. The various modules of the IoT-MP's manager guarantee that the location information of things are only disclosed to authenticated and authorized entities while preserving the location privacy of things and that of their owners. Specifically, the Privacy Module (PM) provides fine-grained features to dynamically adapt the users' defined policies and location information granularity to specific contexts. Currently, we are evaluating, using simulations, the effectiveness and performance of the IoT-MP in preserving the location privacy of the users in a heterogeneous network. Various network simulation studies are planned for future works. The simulations will incorporate the design of an IEEE 802.11ah and ZigBee heterogeneous network. This should provide the research with feedbacks on the performance of the IoT-MP and validate its privacy preserving approach.

**ACKNOWLEDGEMENTS**

This research is supported by the International Postgraduate Research Scholarship (IPRS) and the Australian Postgraduate Award (APA).

**AUTHORS**

**Mahmoud Elkhodr** is with the School of Computing, Engineering and Mathematics at Western Sydney University (Western), Australia. He has been awarded the International Postgraduate Research Scholarship (IPRS) and Australian Postgraduate Award (APA) in 2012-2015. Mahmoud has been awarded the High Achieving Graduate Award in 2011 as well. His research interests include: Internet of Things, e-health, Human Computer-Interactions, Security and Privacy.

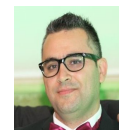

**Dr. Seyed Shahrestani** completed his PhD degree in Electrical and Information Engineering at the University of Sydney. He joined Western Sydney University (Western) in 1999, where he is currently a Senior Lecturer. He is also the head of the Networking, Security and Cloud Research (NSCR) group at Western. His main teaching and research interests include: computer networking, management and security of networked systems, analysis, control and management of complex systems, artificial intelligence applications, and health ICT. He is also highly active in higher degree research training supervision, with successful results. 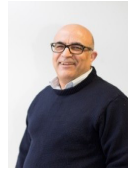

**Dr. Hon Cheung** graduated from The University of Western Australia in 1984 with First Class Honours in Electrical Engineering. He received his PhD degree from the same university in 1988. He was a lecturer in the Department of Electronic Engineering, Hong Kong Polytechnic from 1988 to 1990. From 1990 to 1999, he was a lecturer in Computer Engineering at Edith Cowan University, Western Australia. He has been a senior lecturer in Computing at Western Sydney University since 2000. Dr Cheung has research experience in a number of areas, including conventional methods in artificial intelligence, fuzzy sets, artificial neural networks, digital signal processing, image processing, network security and forensics, and communications and networking. In the area of teaching, Dr Cheung has experience in development and delivery of a relative large number of subjects in computer science, electrical and electronic engineering, computer engineering and networking. 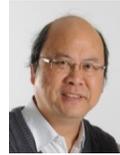